\documentclass[9pt,twocolumn,twoside]{osajnl}
\usepackage{siunitx}
\usepackage{gensymb}
\journal{ol} % Choose journal (ao, aop, josaa, josab, ol, pr)

\setboolean{shortarticle}{true}
\title{Narrow Linewidth near-UV InGaN Laser Diode based on External Cavity Fiber Bragg Grating}

\author[1]{Antoine Congar}
\author[1]{Mathilde Gay}
\author[1]{Georges Perin}
\author[1]{Dominique Mammez}
\author[1]{Jean-Claude Simon}
\author[1]{Pascal Besnard}
\author[2]{Julien Rouvillain}
\author[2]{Thierry Georges}
\author[3]{Laurent Lablonde}
\author[3]{Thierry Robin}
\author[1]{Stéphane Trebaol}

\affil[1]{Univ Rennes, CNRS, Institut FOTON - UMR 6082, F-22305 Lannion, France}
\affil[2]{Oxxius, 4 rue Louis de Broglie, 22300 Lannion}
\affil[3]{iXblue, rue Paul Sabatier, 22300 Lannion}

\affil[*]{Corresponding author: stephane.trebaol@enssat.fr}

\begin{abstract}
We realize a fiber Bragg grating InGaN based laser diode emitting at 400 nm and demonstrate its high coherency. Thanks to the fabrication of a narrow band fiber Bragg grating in the near-UV, we can reach single-mode and single-frequency regimes for the self-injection locked diode.  The device exhibits 44 dB side-mode-suppression-ratio and mW output power. Detailed frequency noise analysis reveals sub-MHz integrated linewidth and 16 kHz intrinsic linewidth.  Such a narrow linewidth laser diode in the near-UV domain with a compact and low-cost design could find applications whenever coherency and interferometric resolutions are needed.
\end{abstract}

\setboolean{displaycopyright}{true}

\begin{document}

\maketitle

\begin{figure}[htbp]
	\centering
	\fbox{\includegraphics[width=0.9\linewidth]{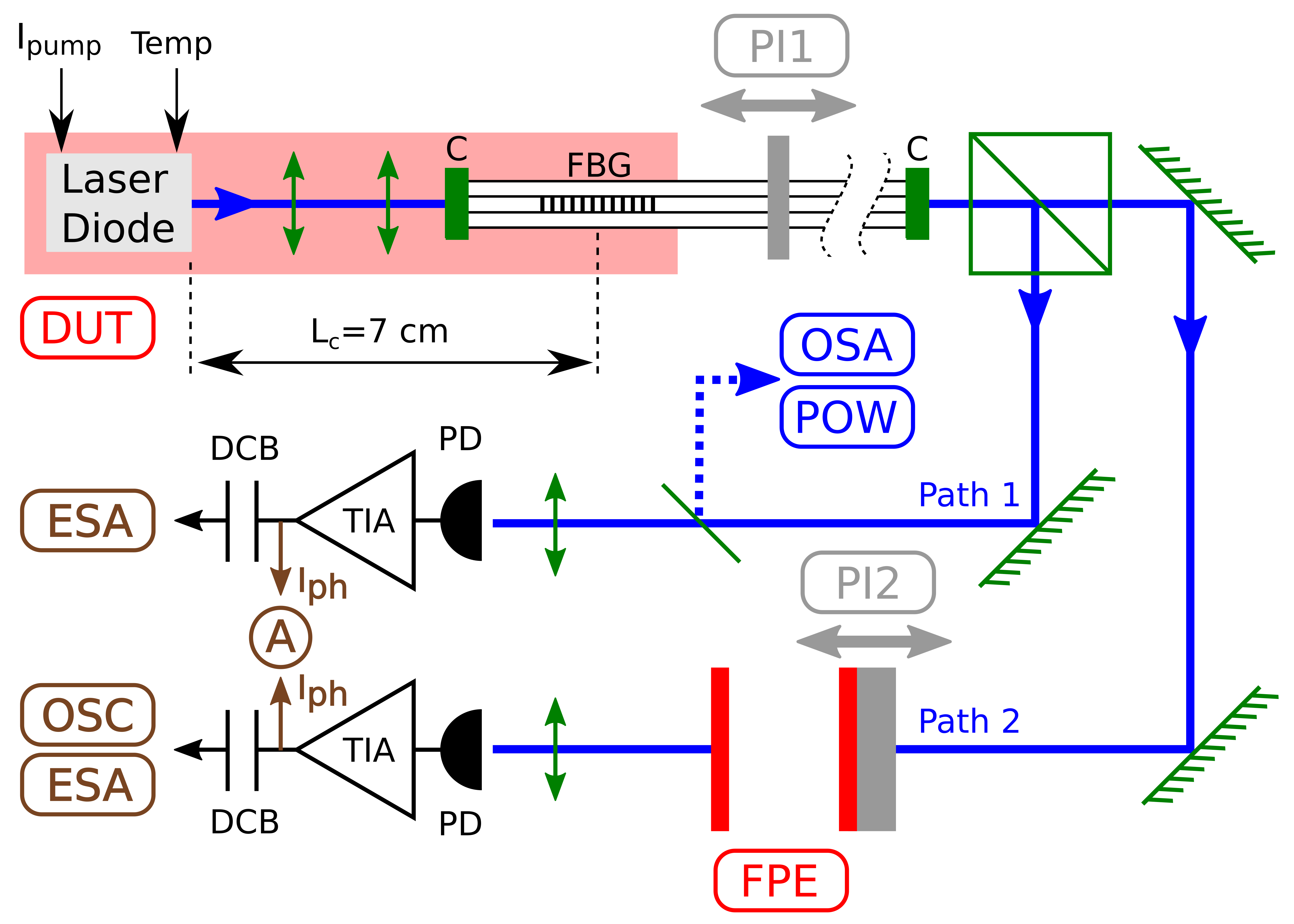}}
	\caption{Experimental setup. The device under test (DUT) is composed of the LD, optical beam shaping lenses and a fiber Bragg grating (FBG). The external cavity length (about 7 cm long) is adjusted using a piezo actuator (PI1). Path 1 is used for intensity noise (IN), low resolution spectrum using an optical spectrum analyzer (OSA) and power measurements (POW). For OSA and POW measurements, a fraction of the signal is extracted. On path 2, the movement of one mirror of the Fabry-Perot etalon (FPE) is used for laser line scanning and frequency noise measurement, thanks to a piezo actuator (PI2). Electrical signals generated by photodiodes (PD) are sent to an oscilloscope (OSC) and an electrical spectrum analyzer (ESA) through transimpedance amplifiers (TIA) and DC-blocks (DCB). DC photocurrents are measured using a multimeter (A). Coupling of light in the fiber and collimation of the output beam are made using APC fiber couplers (C).}
	\label{fig:setup}
\end{figure}

\begin{figure}[htbp]
	\centering
	\fbox{\includegraphics[width=0.9\linewidth]{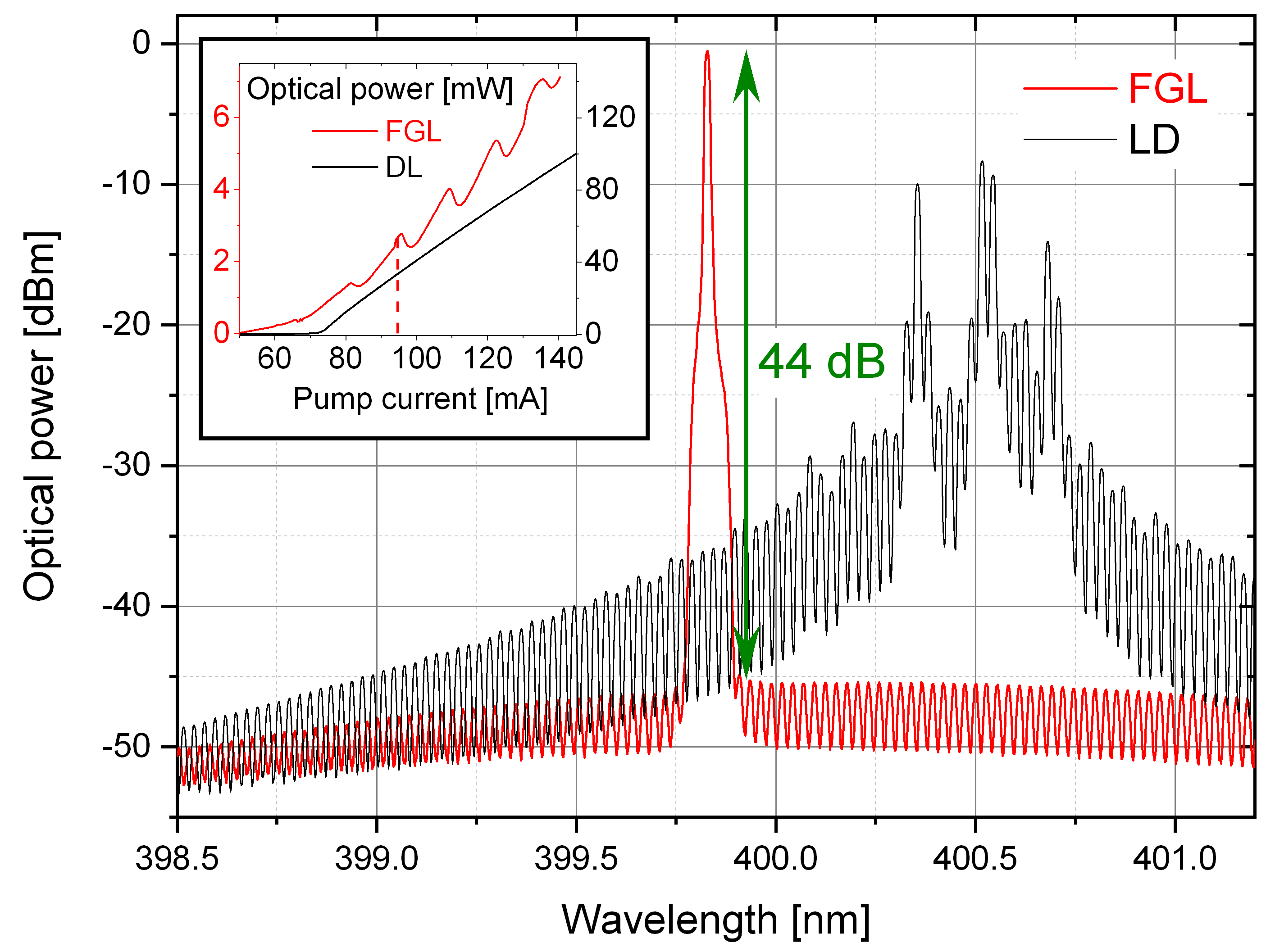}}
	\caption{Spectrum of the solitary laser diode (black curve) and FGL (red curve) for a pump current close to 95 mA. Inset) L-I curves of the solitary laser diode (black curve) and FGL (red curve).}
	\label{fig:spectrum}
\end{figure}

The InGaN-based laser diodes (LD) market is mainly driven by the Blu-ray industry, which requires powerful and low-cost sources. The technology is now mature, providing reliable laser products in the blue/violet range and extends to UV. However, the need for narrow linewidth LDs is growing in a variety of domains ranging from industrial to metrological applications, where linewidth requirements extends widely from tens of GHz to sub-MHz. Compact and low-cost LDs are mandatory to address applications such as visible light communication \cite{Wu2017}, underwater LiDAR sensing \cite{Rumbaugh2016}, 2D and 3D holographic storage \cite{Tanaka2007} and industrial spectroscopy \cite{Larkin2011}. Furthermore, fundamental spectroscopic applications, optical clocks, atom cooling and atom interferometry applications require highly coherent sources with accurate wavelength for the pumping of particular transitions or probing hyperfine atomic structures \cite{Oates1999,Zeng2014}.
Those applications would benefit from low-cost, compact and highly coherent LD. The research on coherence properties of GaN edge-emitting LDs is still in its infancy. Two main designs are considered in the literature to reach single-longitudinal-mode (SLM) regime.
The first design called "monolithic approach" relies on the ridge waveguide effective index modulation. Recently, first electrically pumped SLM laser diode using a high order grating has been demonstrated \cite{Slight2018}. In this work, the authors have reported 35~dB side-mode suppression ratio (SMSR). Despite encouraging results, this approach is not yet mature for mass production and requires demanding technical resources such as high-resolution e-beam lithography.\\
In a second approach, SLM operation is achieved thanks to the filtering function of an external cavity. External-cavity diode lasers (ECDL) are composed of a Fabry-Perot LD and an external mirror providing an optical feedback. Moreover, by transferring its spectral purity to the diode, the use of higher quality factor cavities can induce drastic spectral narrowing enabling single-frequency (SF) operation. These cavities are typically high finesse Fabry-Perot resonators \cite{Horstjann2012} or whispering gallery mode resonators \cite{Donvalkar2018}.
In most commercially available ECDLs, optical feedback is provided by a diffraction grating \cite{Favre1986} whose position and angle with respect to the LD should be accurately controlled. These ECDLs are thus expensive and quite large devices because they require implementation of high-quality electromechanical components and expensive anti-reflection coating LDs to reach their performances.\\
An alternative, called fiber Bragg grating LD (FGL) scheme, has been proposed and extensively studied with applications in the telecom band \cite{Kashyap1999}. Here, the LD is coupled to a narrow band fiber Bragg grating (FBG). The stability lies on the in-fiber optical function and no mechanical component is needed like in conventional ECDL \cite{Ziari97}. The main advantage of FGL devices is their versatility. By design, the Bragg mirror characteristics (center wavelength, peak reflectivity and bandwidth) can be easily and accurately modified so that external cavity parameters can be tailored to optimize SLM operation, where high SMSR and high optical output power are mandatory, or SF operation where narrow linewidth is requested in particular for metrological applications.
In the following, we consider that the SLM operation is obtained when one mode of the laser diode is selected by the Bragg mirror and reflected back in the laser. On top of that, to reach SF operation, a careful tuning of the external cavity length should be done to put in phase one external cavity mode with the previously selected LD mode. Then strong narrowing of the laser emission can be observed. This two regimes address the large range of linewidth satisfying the application needs expressed above.\\
In this paper, we demonstrate near-UV (NUV) FGL exhibiting stable SLM emission at 400~nm with 1.3~mW optical power, 44~dB SMSR and intensity noise (IN) below -130~dB/Hz above 10~kHz. SF operation can even be obtained by accurately setting up the FBG cavity. In this last configuration, we report intrinsic linewidth down to $16 \pm 5$~kHz. In our work a special attention to use low cost components for the laser design is considered. So far, AR-coated NUV laser diodes are dozen times more expensive than simple cleaved facet FP laser diodes. Contrary to what is usually reported in the literature, SLM or SF designs have been obtained by using a simple GaN LD without any AR coating. Our work demonstrates the possibility of achieving state-of-the-art performances by judiciously combining low-cost components.\\
\\
The NUV FGL is schematically represented by a red box on figure \ref{fig:setup}. The laser diode is a commercially available 400 nm emission wavelength InGaN/GaN LD without anti-reflection coating. Cleaved facets determine the laser cavity, which is typically in the order of 1 mm long. The free spectral range (FSR) is measured to be 27 pm (50.5 GHz).  By design, the diode emits on the fundamental transverse mode (TEM00) but the beam is elliptical and highly divergent. A set of lenses is used to circularize and collimate the beam, which is then focused into a single-mode fiber through a fiber coupler (C). 
The external cavity is based on a Bragg mirror, photo-inscribed in the core of a single-mode fiber. To eliminate parasitic reflections, fiber ends are polished at an angle of 8$\degree$ (APC connectors). The overall length of the cavity is 7~cm, including the free space section used for beam-shaping and the few centimeters fiber section extending to the Bragg mirror. The Bragg mirror inscription relies on the photosensitivity of germanosilicate single-mode fiber (core diameter 2.4~$\mu$m at 400~nm). The fiber is transversally exposed to a UV fringe field. No phase-mask is commercially available to reach Bragg wavelengths lower than 405~nm by direct inscription technique. We thus used a Talbot interferometer allowing Bragg mirrors with reflectivity wavelengths within the range 375-405~nm \cite{Stump2000}. To our knowledge, it is the first realization of FBG at such short Bragg wavelengths. \\
To ensure the selection of only one longitudinal mode of the LD cavity, the 3~dB-reflection-bandwidth of the FBG has to be smaller than the FSR. We then designed the FBG bandwidth to be around 20 pm. Moreover, to reach strong feedback regime \cite{Petermann1988}, the following expression should be satisfied : $I^2 R_{Bragg}>R_{LD}$, where $I=0.7$ is the fiber coupling coefficient, $R_{LD}=0.18$ the output LD cavity reflectivity mirror and $R_{Bragg}$ the Bragg mirror reflectivity. Coefficient $R_{Bragg}$ should thus be greater than 0.4. Hence, the choice of a low-cost non-AR coated LD implies the use of a quite high reflectivity Bragg mirror. Thus, the peak reflectivity of the Bragg mirror is chosen to be $R_{Bragg}=70\%$.\\
L-I curves of the laser diode without (black curve) and with (red curve) optical injection from the FBG are shown in the inset of figure \ref{fig:spectrum}. As expected, the current threshold is reduced from 73 to 50 mA. Under optical feedback, the L-I curve displays a periodic modulation of the power that corresponds to mode hopping from adjacent LD modes. The highly multi-longitudinal mode solitary LD spectrum is shown in black line on figure \ref{fig:spectrum}.
The central wavelength is close to 400.5~nm and the spectrum spreads over $\simeq$4~nm (at 40~dB from the maximum). Spectral structuration can be seen, resulting from a phenomenon, well known in GaN-based LDs, called mode clustering \cite{Congar2017}.\\
%When feedback is provided, SLM regime is obtained with 1.3~mW optical power. The linewidth is lower than 10~pm, which corresponds to the resolution of the OSA and an SMSR higher than 44 dB is obtained (red curve on figure \ref{fig:spectrum}). To reach such a SLM regime, the Bragg resonance should match with one particular LD mode. This matching is obtained by tuning the pump current and the Bragg wavelength to improve overlap between a specific LD mode and the Bragg central wavelength. Furthermore, this wavelength should be as close as possible to the LD gain curve maximum, in order to maximize the SMSR.
When adding the FBG with a bandwidth smaller than the laser diode FSR, one single mode is selected as it can be seen on figure \ref{fig:spectrum}. To reach such a SLM regime, the Bragg resonance is adjusted to one particular LD mode by tuning the pump current and the Bragg wavelength to improve overlap between a speciﬁc LD mode and the Bragg central wavelength. Furthermore, this wavelength should be as close as possible to the LD gain curve maximum, in order to maximize the SMSR. In our experimental conditions, we obtained an SMSR of 44 dB and an optical power of 1.3 mW as depicted in figure \ref{fig:spectrum}. It is to notice that the fine structure of the spectrum is not resolved by the 10pm resolution OSA used and characterisations are completed by further measurements as described in the following.
Comparable SMSR performance, close to 40 dB, has been obtained with a bulk diffraction grating configuration in reference \cite{Ruhnke2014}. A value of 25~dB has been measured for an equivalent structure without AR coating in the telecom range \cite{Park1986}. It is to notice that best results for SLM lasers obtained by the monolithic approach at NUV wavelengths are 25~dB SMSR \cite{Slight2018}.
Futhermore, FBG stretching can be used to tune the SLM central frequency over 0.5 nm with mode hops between adjacent LD longitudinal modes separated from an FSR.
To this aim, one FBG extremity is fixed, and the other is attached on a translation plate connected to a piezo actuator. Applying a longitudinal strain onto the fiber shifts the FBG wavelength with a sensitivity of 0.3~pm/µstrain.
This configuration provides stable SLM operation over a period of hours. The device fulfills requirements for SLM applications.
\\\\
To get single-frequency (SF) operation, the external cavity length must be precisely
controlled in order to reach good spectral overlap between a mode of the LD cavity and a mode of the external cavity.
\begin{figure}[htbp]
	\centering
	\fbox{\includegraphics[width=0.9\linewidth]{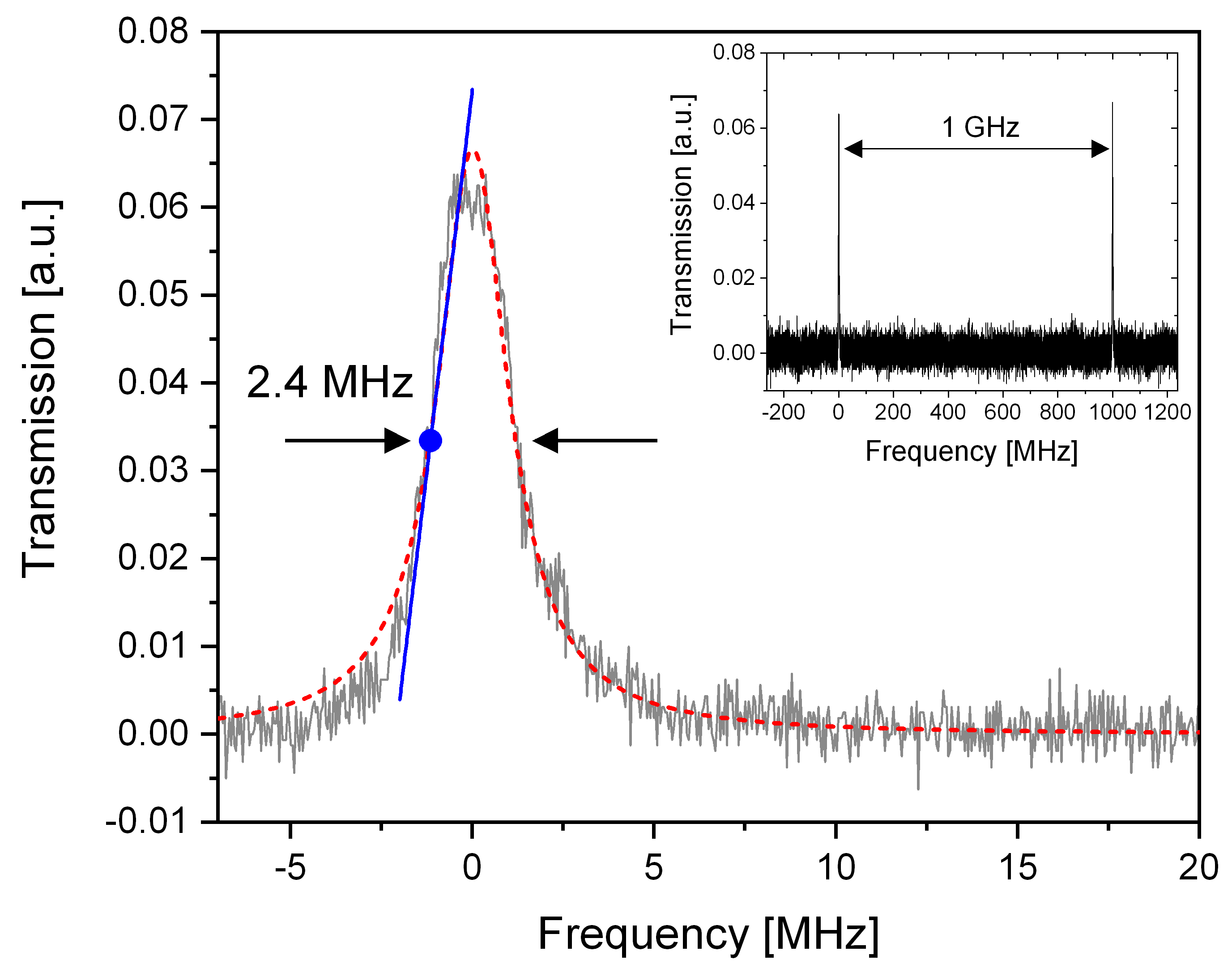}}
	\caption{FGL linewidth estimation obtained by characterizing the Fabry-Perot etalon (FPE) transmission for 63 mA of pump current. The red dashed line is a Lorentzian fit of the measurement. One FPE FSR is shown in the inset to ensure single-frequency operation and for frequency graduation of the abscissa axis. During FN measurement, the FPE, used as a frequency discriminator, is maintained at the quadrature point (blue point) where response is linear (blue linear fit).}
	\label{fig:FP}
\end{figure}
Figure \ref{fig:setup} displays the linewidth and frequency noise measurement benches we used. The laser fibered output is separated into two paths. Path 1 is used to measure intensity noise, optical power and low-resolution spectrum while path 2 gives access to frequency noise and real-time high-resolution spectrum measurement. For noise measurements (IN and FN), the signal is focused on a photodiode and measured as a power spectral density (PSD) using an electrical spectrum analyzer (ESA) through a variable gain transimpedance amplifier.\\
High resolution spectrum measurement is performed using a Fabry-Perot etalon (FPE) (FSR 1GHz, finesse>500) on path 2. A FPE resonance is scanned over the laser line, applying a voltage ramp to a piezoelectric actuator and the output signal is observed as a function of time on an oscilloscope.

Measurement of one FSR of the 1~GHz FP etalon (inset in figure \ref{fig:FP}) is used to ensure single-frequency operation and scale the abscissa axis as a function of frequency. From a close look on the laser line shown on figure \ref{fig:FP}, the linewidth is estimated to be $\Delta\nu$=2.4~MHz, which corresponds to the FPE resolution.
\\We then performed FN characterization to get more insights on the SF laser frequency dynamics and obtain an estimation of the laser linewidth for selected integration times. The FP voltage ramp is removed and the moving mirror is now connected to the output of a PID controller, which parameters are set in such a way that the laser is maintained at the quadrature point (blue dot in figure \ref{fig:FP}). In such a configuration, intensity fluctuations of the output signal are proportional to frequency fluctuations of the input signal and the proportionality coefficient is given by the resonance flank slope ($P=35 \pm 5$~mV/MHz). By measuring the PSD of the output signal, we can extract the frequency noise as a function of the Fourier frequency \cite{Fritschel1989}. Intensity noise of the FGL are included in this measurement. IN measurement on path 1 is thus used to guarantee that the intensity noise PSD is negligible in comparison with the FN PSD. Indeed, IN PSD curve is at least 20~dB below the FN PSD in the frequency range of interest. We measure the relative intensity noise of the laser to be -130~dB/Hz above 10~kHz.
\begin{figure}[htbp]
	\centering
	\fbox{\includegraphics[width=0.9\linewidth]{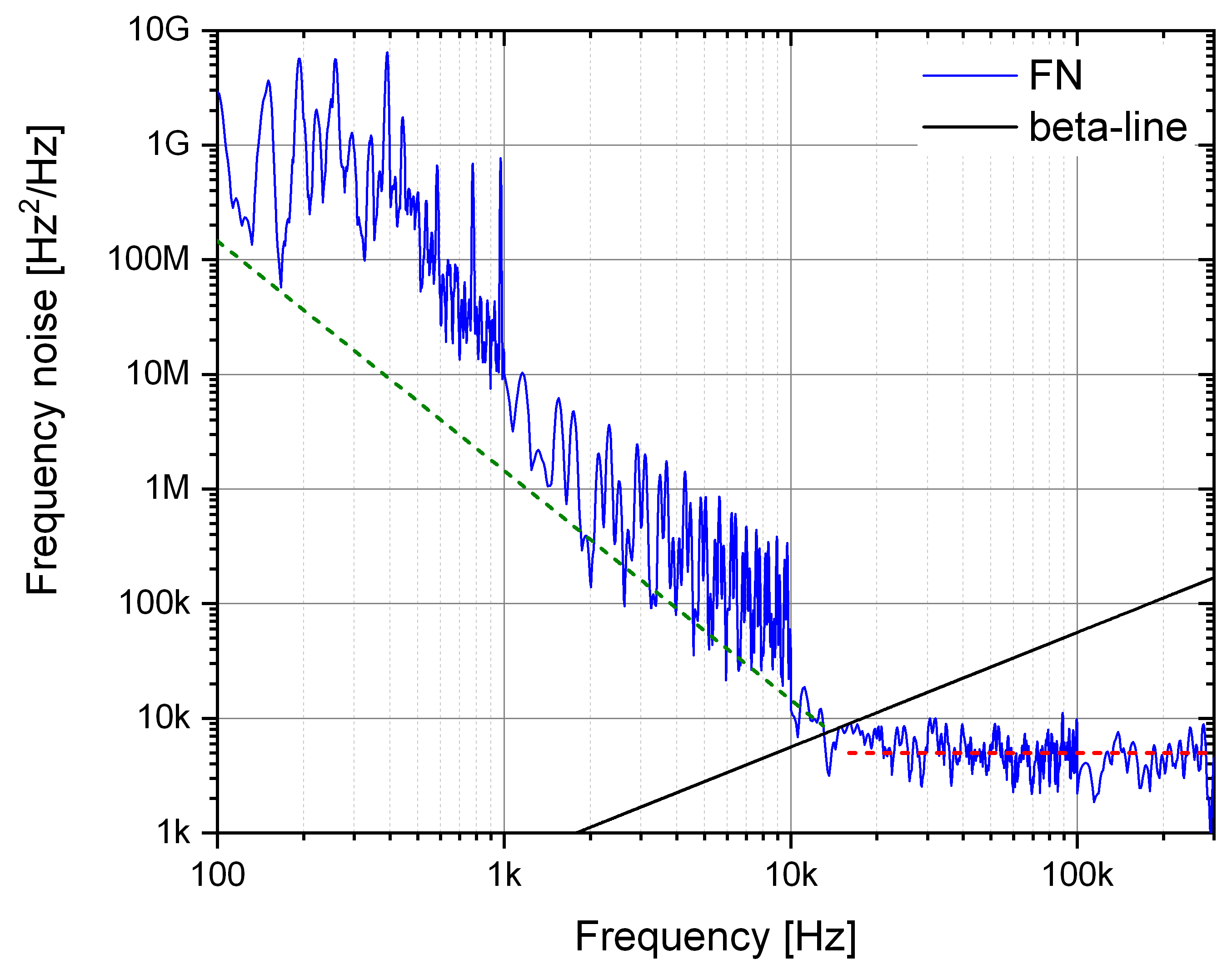}}
	\caption{Frequency noise measurement. Peaks in the curve are due to external perturbations. The green dashed line shows the low frequency $1/f^2$ trend, the red dashed line the high frequency plateau. The servolocking bandwidth is limited to 20~Hz to insure no parasitic contributions to the FN measurement.}
	\label{fig:FN}
\end{figure}
\\In a laser, frequency noise can usually be described by two main contributions (see Fig. \ref{fig:FN}) expressed by $S_{\nu}=\frac{h_{\alpha}}{f^{\alpha}}+h_0$. A high frequency white noise, which is related to the Lorentzian shape of the intrinsic laser-linewidth (dashed red line). This white noise level is identified by the coefficient $h_0$. This noise is due to the random phase fluctuation of spontaneous emission \cite{Agrawal}.
%[p. 269]
At lower frequencies the device is under influences of various contributions like acoustic, thermic and electromagnetic perturbations. Their spectral signature usually follows a $f^{-\alpha}$ evolution on the frequency noise with $\alpha$ bounded between 1 and 2. This second contribution gives a Gaussian shape to the integrated optical linewidth \cite{DiDomenico2010} that exhibits a Voigt profile in the general case \cite{Stephan2005}.\\ 
FN measurement plotted in figure \ref{fig:FN} is taken at pump current I=63~mA. We find the two behaviors mentioned above. At low frequency (<10~kHz), the FGL FN displays a $1/f^2$ tendency with  $h_2=1.45\times 10^{12}$~Hz$^2$/Hz (green dashed line in the figure). Strong perturbations probably coming from the current source lead to an increase of FN beyond the $1/f^2$ line (100~Hz-1~kHz). Above 10~kHz, the FN reaches a plateau at $h_0=5\times 10^3$~Hz$^2$/Hz, corresponding to the white intrinsic noise of the laser.\\ 
We use the beta-line approach \cite{DiDomenico2010} to estimate the integrated optical linewidth of the device. Short-time (t=1ms) and longer-time (t=10ms) linewidths are 250~kHz and 950~kHz respectively. In this calculation, peaks from external perturbations have been neglected. The former value is in the same order of magnitude than the linewidth of 420 nm grating-ECDL laser presented in \cite{Zeng2014} and 405 nm self-injection locked LD in \cite{Donvalkar2018}. Using the relation $\Delta\nu=\pi h_0$ \cite{DiDomenico2010}, we can estimate the intrinsic linewidth of the laser to be $16 \pm 5$ kHz. Savchenkov \emph{et al.} \cite{Savchenkov2019} have reported a similar value ($\approx$ 30 kHz) for self-injection locked LD at 370 nm. ECDL frequency stabilizations based on external cavity optical feedback \cite{Horstjann2012,samutpraphoot2014} allow to narrow the linewidth down to few kHz but the large external cavity (meter long) and moving parts limit the observation time from minutes to seconds.\\
%ECDLs with narrower linewidths, below 1~kHz, can be obtained \cite{Wyatt85, Petermann1988}.
%%Petermann p.264]
%These performances require high feedback coefficient which can be provided either by the use of a long cavity or a strong feedback cavity \cite{Petermann1988}. Narrowing effect should thus be provided by a strong feedback amount, that is $I^2 R_{Bragg}>R_{LD}$, where $I=0.7$ is the fiber coupling coefficient. Coefficient $R_{Bragg}$ should thus be greater than 0.4. Hence, the choice of a low-cost non-AR coated LD implies the use of a quite high reflectivity Bragg mirror.\\ 
To reach stable SF operation, a careful consideration should be given to the setup. External perturbations upon the external cavity are indeed detrimental to the device linewidth performances. Acoustic perturbations are drastically reduced using an anti-vibration table and a Peltier module situated below the external cavity that provides a precise temperature stabilization. However, for the sake of compactness, improvements can be implemented. To ensure stability of the device under single-frequency operation, the amount of reinjected light into the LD after a round trip in the cavity must be as constant as possible. Because LD emission is linearly polarized, the use of polarization-maintaining fiber during the light round trip may ensure higher performances without demanding implementation of other external parts. The cavity length should be reduced as much as possible to decrease external perturbations on the fiber. However, a trade-off has to be found between the length and the quality factor of this cavity \cite{Naumenko2003}.\\
GaN based LDs still suffer from epitaxial growth imperfections that revealed nonlinear gain behaviors \cite{Congar2017}. Moreover, working in the strong feedback regime contributes to favor the appearance of nonlinear effects since a relatively large amount of light is reinjected in the laser diode. To observe SF regime, we thus operate the laser device at low power to prevent the apparition of unstable regimes. In a future work, optimization of the Bragg reflectivity should allow to reach higher power while maintaining SLM or SF regimes.\\
\\
We demonstrate a reliable, all-fibered optical output, compact, FBG NUV laser source. We have shown that single longitudinal mode operation can be achieved using non-antireflection coated GaN LD coupled to a fiber Bragg grating designed at 400~nm. We measured side-mode suppression ratio up to 44~dB with 1.3~mW output power and an intensity noise level below -130~dB/Hz above 10~kHz. Furthermore, transferring the spectral purity of the external cavity to the diode leads to single-frequency operation with integrated sub-MHz linewidth and $16 \pm 5$~kHz intrinsic linewidth. Hence, NUV FGLs could constitute a commercial alternative to expensive grating ECDLs with similar performances. For these reasons, they have been used for decades in various applications at telecom wavelengths where low–cost and compactness have to be associated to high coherency performances. Through this work, we have highlighted that they could also be assets for applications in the NUV.

\section*{Funding}
The present work is supported under projects DeepBlue and UV4Life by the Region Bretagne (contracts N° 16008022, 19005486) and the European Regional Development Fund (contracts N° EU000181, EU000998).

\section*{Disclosures}
The authors declare no conflicts of interest.

% Bibliography
\bibliography{biblio}
% Full bibliography added automatically for Optics Letters submissions; the following line will simply be ignored if submitting to other journals.
% Note that this extra page will not count against page length
%\bibliographyfullrefs{biblio}
\ifthenelse{\equal{\journalref}{ol}}{
\clearpage
\bibliographyfullrefs{biblio}
}{true}

\end{document}